# Numerical Stability Issues and a Mitigation Method in Switching Events Calculation Considering State Variable Discontinuities for EMT Simulation

Sheng Lei
Hitachi Energy
Santa Clara, CA, USA
sheng.lei@hitachienergy.com

*Abstract*—**Some existing methods for switching events calculation considering state variable discontinuities for EMT simulation can encounter numerical stability issues, either over-stabilizing or destabilizing the studied power systems. This paper documents these issues in detail with a concrete example. The underlying mechanism of the issues are revealed via systematic analysis. A mitigation method for the issues is proposed. The convergence behavior of the method is analyzed. Numerical examples demonstrate the effectiveness of the method. It is observed that the method also helps to improve the accuracy of the simulation results.**

## I. Introduction

Power system electromagnetic transient (EMT) simulation usually adopts the implicit trapezoidal method (TR) given a user-defined fixed step size to discretize the differential equations for the regular forward time stepping [1]-[2]. The discretized models of network elements are converted into Norton equivalent circuits and organized via Nodal Analysis [1]-[2]. If switching events (e.g., tripping/connecting a device, turning on/off a power electronic valve) occur, linear interpolation is usually utilized to locate the switching time and bring the system values back to that time instant [1]-[7]. Once the switching events calculation has been settled, the simulation restarts as forward time stepping. Beyond the usual practice, some EMT simulators may adopt other numerical integrators such as the root-matching method [2] and TR-BDF2 [8]. Similarly, quadratic interpolation may be utilized in place of linear interpolation [8].

A major task of switching events calculation is to resolve simultaneous switching, namely, to determine the final switch statuses and system values at the switching time instant after all the required switching events have taken place. Simultaneous switching refers to the phenomena that one or more switching events trigger a series of other causally related switching events at the same time instant [5]-[7], [9]. Potential Dirac impulses play a key role in simultaneous switching determining the switch statuses, therefore differential state variables (e.g., inductor currents, capacitor voltages) should never be assumed continuous [3], [9]-[10]. After resolution of simultaneous switching, reinitialization is usually implemented to obtain reasonably accurate system values for the coming forward time stepping and to suppress numerical oscillations induced by TR [5]-[6], [9]. Alternatively, the Critical Damping Adjustment (CDA) can be implemented, which performs two half time steps of the backward Euler method (BE) and restarts the simulation one full time step after the switching time instant [2], [11]. Obviously, BE deteriorates the accuracy of TR. More severely, switching events within this time step are ignored, if any.

Methods involved in switching events calculation have been proposed in the literature [1]-[7], [9]-[10]. Among them, some building blocks can be recognized: A) a half time step of BE, B) two half time steps of BE and an intermediate linear extrapolation, C) two consecutive half time steps of BE and a linear extrapolation. Note that in EMT simulation, BE is commonly paired with a half time step, because it results in the same nodal conductance matrix as that resulting from TR with a full time step, which saves computations [2], [11]. Resolution of simultaneous switching consists of repetitive executions of one of the building blocks, and switch status checking and changing [3], [5]-[7], [9]-[10]. In reinitialization, on the other hand, the building block is executed only once [5]-[6], [9]. Three methods for accurate calculation of switching events considering state variable discontinuities are proposed and examined in [9]. There, Methods A, B and C are based on the above-mentioned building blocks A, B and C respectively. The same nomenclature will be adopted in this paper.

Further investigations on the three methods in [9] show that they can encounter numerical stability issues especially when applied to power systems with high switching frequencies. As a follow-up study based on [9], this paper documents the issues, analyzes the mechanism, and puts forward a mitigation method. Main contributions of the current paper are twofold. First, the mechanism of the issues is revealed via systematic analysis. Second, a mitigation method is proposed; convergence analysis of the method is provided as well. The effectiveness of the proposed mitigation method is demonstrated through numerical examples, which also show that it helps to improve simulation accuracy.

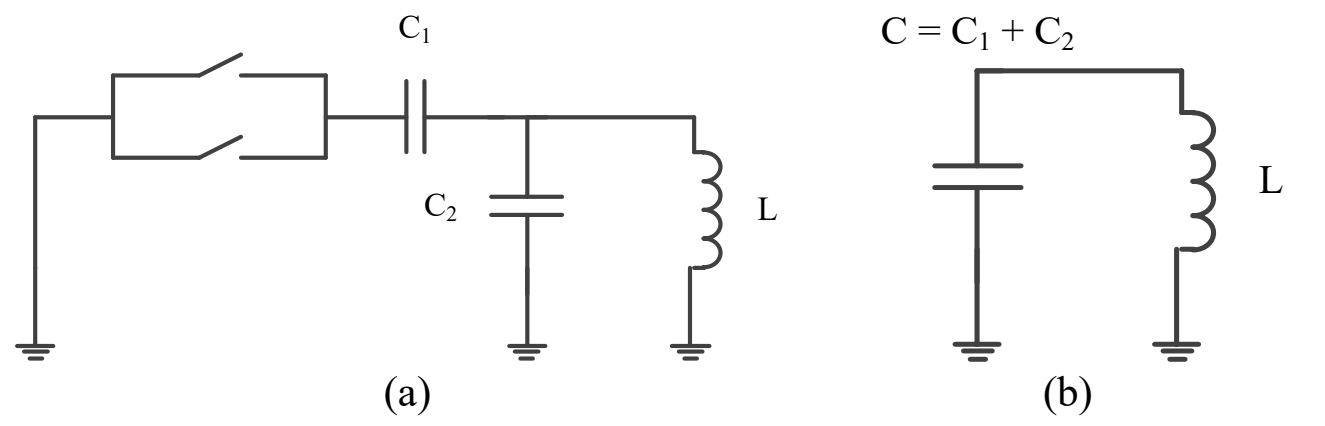

Figure 1. A circuit with switches and the equivalent circuit.

The remainder of this paper is organized as follows. Section II presents the numerical stability issues with the three methods using a concrete example. Section III analyzes the mechanism of the issues. The mitigation method is proposed in Section IV, where its convergence behavior is also analyzed. Numerical examples test the proposed mitigation method in Section V. Finally, Section VI concludes the paper and points out some directions for future research.

## II. Numerical Stability Issues

Consider the circuit shown in Fig. 1 (a), which consists of two ideal switches, two capacitors and an inductor. The switches are adopted to introduce artificial switching events. Whenever the upper switch is closed, the lower switch will be opened, and vice versa. The frequency of the switching events is 20 kHz. In each period, the upper switch is closed for the $1^{st}$ half, and then the lower switch. Clearly, the circuit is equivalent to a simple parallel LC circuit as shown in Fig. 1 (b), especially if the capacitor voltage and the inductor current are of interest.

The equivalent circuit can be described mathematically as

$$\frac{d}{dt}\begin{pmatrix} v_C \\ i_L \end{pmatrix} = \begin{pmatrix} 0 & -\frac{1}{C} \\ \frac{1}{L} & 0 \end{pmatrix}\begin{pmatrix} v_C \\ i_L \end{pmatrix} \tag{1}$$

where t is the time; $v_C$ is the capacitor voltage; $i_L$ is the inductor current. The circuit is marginally stable, having eigenvalues at $\pm i\,{}^{1}\!/\!\sqrt{LC}$. Given initial condition $v_C(t_0)$ and $i_L(t_0)$, the analytical solutions can be calculated as

$$\begin{cases} v_C(t) = v_C(t_0)\cos(\frac{1}{\sqrt{LC}}(t-t_0)) - i_L(t_0)\sqrt{\frac{L}{C}}\sin(\frac{1}{\sqrt{LC}}(t-t_0)) \\ i_L(t) = i_L(t_0)\cos(\frac{1}{\sqrt{LC}}(t-t_0)) + v_C(t_0)\sqrt{\frac{C}{L}}\sin(\frac{1}{\sqrt{LC}}(t-t_0)) \end{cases} \tag{2}$$

where $t_0$ is the start time. Clearly, both the capacitor voltage and the inductor current are purely sinusoidal.

In the following simulations, parameters and variables are expressed in per-unit values except for the time. $C_1 = 4 \times 10^{-4}$, $C_2 = 2 \times 10^{-4}$, $L = 2 \times 10^{-4}$, $v_C(t_0) = 0.5$, $i_L(t_0) = 1.0$. A fixed step size of 20 $\mu s$ is adopted for the regular forward time stepping, which is equally divided into two sub-time steps to enable quadratic interpolation for switching event detection and location. This setting brings higher accuracy, helping to focus the study attention on simultaneous switching resolution and reinitialization. Methods A, B and C for switching events calculation are applied to the circuit with switches combined with the forward time stepping respectively. Numerical results obtained with these methods are compared to the analytical solutions in Fig. 2.

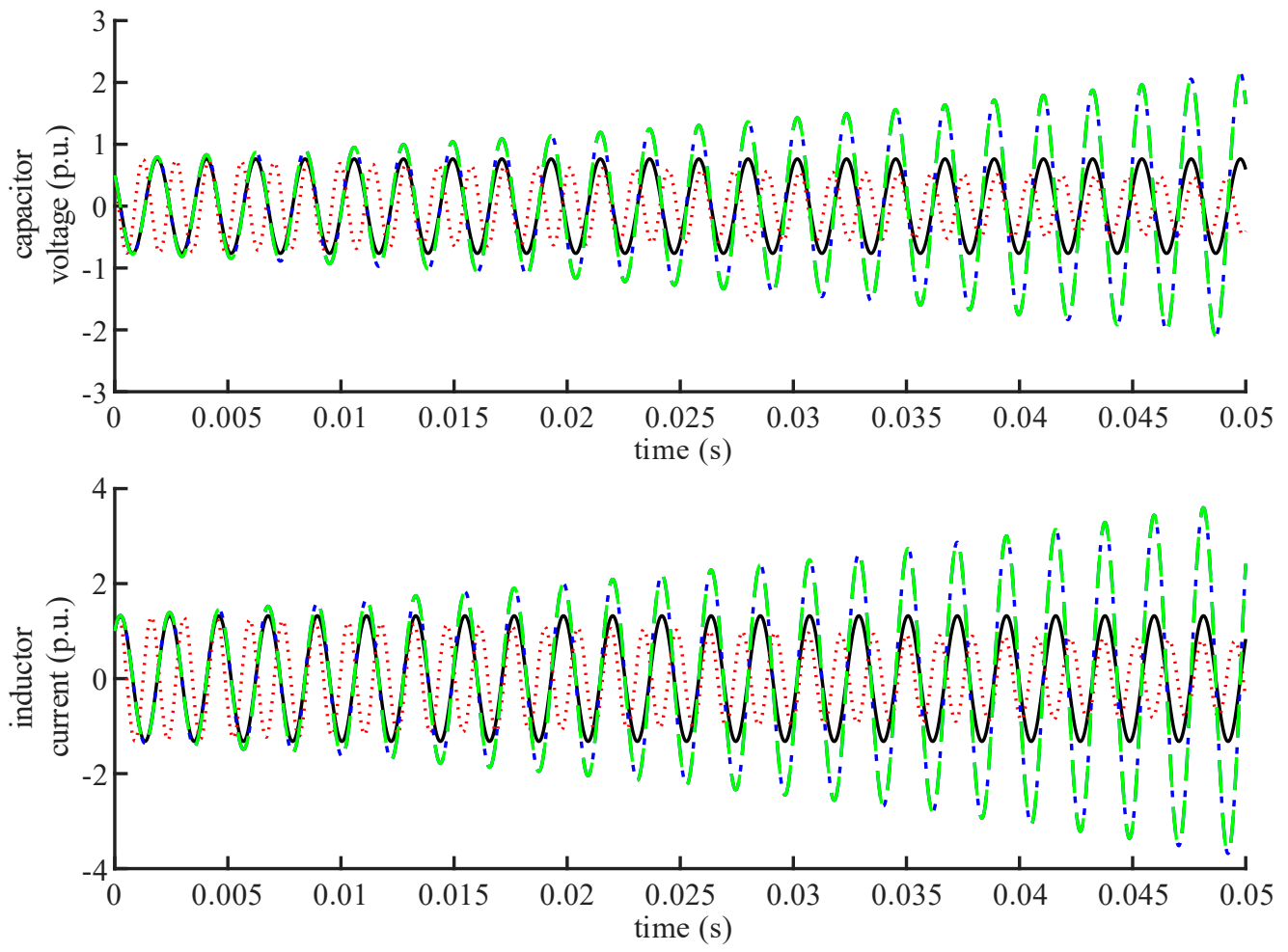

Figure 2. Results from different methods for switing events calculation compared to the analytical solutions. Solid line: analytical. Dotted line: Method A. Dash-dotted line: Method B. Dashed line: Method C. The dash-dotted line and dashed line basically overlap each other in both subplots.

From Fig. 2, it can be observed that Method A introduces obvious numerical damping, overly stabilizing the circuit. Methods B and C make the circuit numerically unstable. The results of Methods B and C are basically the same, which is also reported in [9]. All the three methods have their own numerical stability issue.

## III. Mechanism Analysis

Numerical stability property of the three methods is to be analyzed in this section using a technique similar to that in stability analysis of numerical integrators. In this paper, $t_{SW}$ denotes a generic switching time instant; $t_{SW-}$ denotes the time instant immediately before; $t_{SW+}$ denotes the time instant immediately after.

### *A. Method A: a Half Time Step of BE*

In Method A, BE with a half time step is adopted in the simultaneous switching resolution phase to calculate $t_{SW}$ from $t_{SW-}$. In the s-domain, the calculation is

$$X_{t_{SW}} = X_{t_{SW-}} + \frac{h}{2} s X_{t_{SW}} \tag{3}$$

where X is a generic differential state variable in the s-domain; h is the step size; s is the differential operator in the s-domain. (3) can be reformulated as

$$X_{t_{SW}} = \frac{1}{1-\frac{h}{2}s} X_{t_{SW-}} \tag{4}$$

Then BE is adopted in the reinitialization phase to calculate $t_{SW+}$ from $t_{SW}$ as

$$X_{t_{SW+}} = \frac{1}{1-\frac{h}{2}s} X_{t_{SW}} \tag{5}$$

Therefore, the transfer function of Method A is

$$\left(\frac{1}{1-\frac{h}{2}s}\right)^2 \tag{6}$$

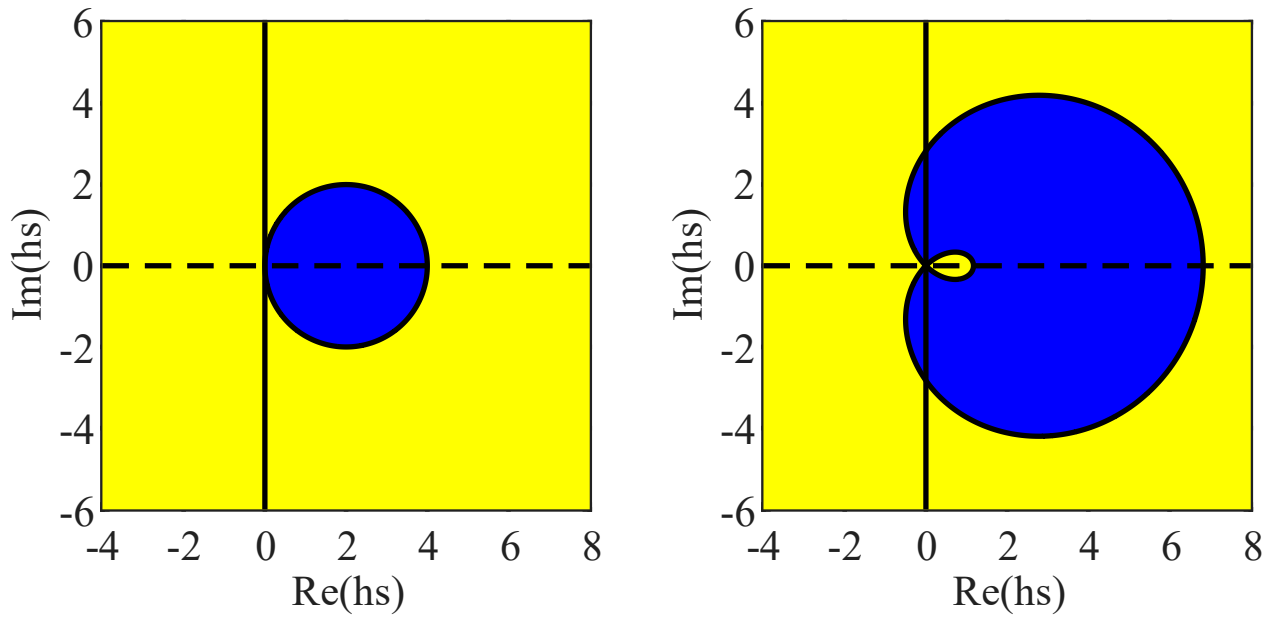

Figure 3. Transfer functions of the Methods. Light-colored area: stable region. Dark-colored area: unstable region.

The transfer function (6) is plotted in Fig. 3 (a), where hs as a whole is considered as the variable. It can be learned that a major part of the right-half plane falls inside the stable region, meaning that Method A can make many physically unstable systems numerically stable, resulting in misleading simulation results. Therefore, we see overly damped waveforms from Method A in Section II.

### *B. Method B: Two Half Time Steps of BE and an Intermediate Linear Extrapolation*

In the simultaneous switching resolution phase, Method B first calculates $t_{SW}+\frac{h}{2}$ from $t_{SW-}$ using BE

$$X_{t_{SW}+\frac{h}{2}} = \frac{1}{1-\frac{h}{2}s} X_{t_{SW-}} \tag{7}$$

Then it calculates $t_{SW}-\frac{h}{2}$ via linear extrapolation based on $t_{SW-}$ and $t_{SW}+\frac{h}{2}$

$$X_{t_{SW}-\frac{h}{2}} = 2X_{t_{SW-}} - X_{t_{SW}+\frac{h}{2}} = \frac{1-hs}{1-\frac{h}{2}s} X_{t_{SW-}} \tag{8}$$

Then it calculates $t_{SW}$ from $t_{SW}-\frac{h}{2}$ using BE

$$X_{t_{SW}} = \frac{1}{1-\frac{h}{2}s} X_{t_{SW}-\frac{h}{2}} = \frac{1-hs}{\left(1-\frac{h}{2}s\right)^2} X_{t_{SW-}} \tag{9}$$

In the reinitialization phase, Method B follows the same procedure, but calculates $t_{SW+}$ from $t_{SW}$

$$X_{t_{SW+}} = \frac{1-hs}{\left(1-\frac{h}{2}s\right)^2} X_{t_{SW}} \tag{10}$$

Therefore, the transfer function of Method B is

$$\left(\frac{1-hs}{\left(1-\frac{h}{2}s\right)^2}\right)^2 \tag{11}$$

The transfer function (11) is plotted in Fig. 3 (b). It can be observed that the unstable region invades parts of the left-half plane, indicating that Method B can turn some physically stable systems into numerically unstable ones, leading to unreliable simulation results. This is the reason for those increasingly enlarged waveforms obtained with Method B in Section II.

### *C. Method C: Two Consecutive Half Time Steps of BE and a Linear Extrapolation*

In the simultaneous switching resolution phase, Method C calculates $t_{SW}+\frac{h}{2}$ as (7), and calculates $t_{SW}+h$ using BE from $t_{SW}+\frac{h}{2}$

$$X_{t_{SW}+h} = \frac{1}{1-\frac{h}{2}s} X_{t_{SW}+\frac{h}{2}} = \frac{1}{\left(1-\frac{h}{2}s\right)^2} X_{t_{SW-}} \tag{12}$$

Then it calculates $t_{SW}$ via linear extrapolation based on $t_{SW}+\frac{h}{2}$ and $t_{SW}+h$

$$X_{t_{SW}} = 2X_{t_{SW}+\frac{h}{2}} - X_{t_{SW}+h} = \frac{1-hs}{\left(1-\frac{h}{2}s\right)^2} X_{t_{SW-}} \tag{13}$$

In the reinitialization phase, Method C follows the same procedure, but calculates $t_{SW+}$ from $t_{SW}$

$$X_{t_{SW+}} = \frac{1-hs}{\left(1-\frac{h}{2}s\right)^2} X_{t_{SW}} \tag{14}$$

Comparing (13) and (9), (14) and (10), it is clear that the transfer function of Method C is the same as Method B.

As they have the same transfer function, Methods B and C share the same numerical stability property. Fig. 3 (b) readily applies to Method C as well.

## IV. Mitigation Method

Aiming at mitigating or even eliminating the numerical stability issues in switching events calculation, a natural idea is to obtain precise values at $t_{SW}$ and $t_{SW+}$. To achieve such precise calculations, a method is proposed in this section.

Specifically, it is suggested that each round of simultaneous switching resolution and reinitialization be respectively split into an estimation stage followed by a correction stage. Method A, B or C is still adopted in the estimation stages. However, the outcomes are not directly accepted; but considered as intermediate estimations that are to be corrected.

### *A. Instantaneous Correction in Simultaneous Switching Resolution*

In simultaneous switching resolution, values at $t_{SW-}$ have been given. Suppose that the derivative of a differential state variable at $t_{SW}$ has been estimated as $\dot{x}^{(0)}(t_{SW})$, it is then used as the starting point for the following iterative correction process

$$x^{(k+1)}(t_{SW}) = x(t_{SW-}) - \frac{h}{2}\dot{x}^{(k)}(t_{SW}) + \frac{h}{2}\dot{x}^{(k+1)}(t_{SW}) \tag{15}$$

The correction process is executed until the difference between two consecutive iterations $x^{(k+1)}(t_{SW}) - x^{(k)}(t_{SW})$ is sufficiently small. In this paper, the correction process is called instantaneous correction. Note that instantaneous correction is different from TR or "Instantaneous Solution" [4].

### B. Convergence Analysis

If a differential state variable is continuous, the calculated value at $t_{SW}$ through instantaneous correction should be almost the same as the given value at $t_{SW-}$; and the intermediate results should converge to the value at $t_{SW-}$.

Suppose that both the differential state variable and its derivative are free. Rewrite (15) in the s-domain

$$X^{(k+1)}(t_{SW}) = X(t_{SW-}) - \frac{h}{2} sX^{(k)}(t_{SW}) + \frac{h}{2} sX^{(k+1)}(t_{SW}) \qquad (16)$$

Equivalently

$$\begin{aligned} &X^{(k+1)}(t_{SW}) - X(t_{SW-}) = \\ &-\frac{h}{2} sX^{(k)}(t_{SW}) + \frac{h}{2} sX^{(k+1)}(t_{SW}) - \frac{h}{2} sX(t_{SW-}) + \frac{h}{2} sX(t_{SW-}) \end{aligned} \qquad (17)$$

After algebraic manipulations

$$X^{(k+1)}(t_{SW}) - X(t_{SW-}) = -\left( \frac{h}{2}s \Big/ 1-\frac{h}{2}s \right)(X^{(k)}(t_{SW}) - X(t_{SW-})) \qquad (18)$$

In order for (18) to be contractive, it is required that

$$\left| \frac{h}{2}s \Big/ 1-\frac{h}{2}s \right| < 1 \qquad (19)$$

$$\Leftarrow \quad \left|\frac{h}{2}s\right|^2 < \left|1-\frac{h}{2}s\right|^2 \qquad (20)$$

$$\Leftarrow \quad \left(\frac{h}{2}s\right)\left(\overline{\frac{h}{2}s}\right) < \left(1-\frac{h}{2}s\right)\left(\overline{1-\frac{h}{2}s}\right) \qquad (21)$$

Note that h is a positive real number

$$\Leftarrow \quad \mathrm{Re}(s) < \frac{1}{h} \qquad (22)$$

Therefore, if applied to a physically stable system, including a marginally stable one, (15) is convergent despite the step size. If applied to a physically unstable system, (15) is convergent given sufficiently small step size.

Next, suppose that the derivative is constrained. For example, when an inductor is connected to a voltage source, the inductor current is continuous, but the inductor voltage becomes the same as the voltage source instantaneously. At the 1st iteration, the derivative will be corrected to that specified value due to the constraint. At the 2nd iteration, (15) becomes

$$x^{(2)}(t_{SW}) = x(t_{SW-}) - \frac{h}{2}\dot{x}(t_{SW}) + \frac{h}{2}\dot{x}(t_{SW}) = x(t_{SW-}) \qquad (23)$$

In this situation, (15) converges in two iterations. Therefore, considering constrained derivatives, instantaneous correction should be executed for at least two iterations.

### C. Discontinuous Differential State Variables

If a differential state variable is not free, but constrained by extra conditions, its value could change instantaneously. For example, when a capacitor with nonzero initial voltage is connected to a voltage source, the capacitor voltage will become the same as the voltage source instantaneously; and the capacitor current will exhibit a Dirac impulse.

Instantaneous correction allows discontinuous differential state variables. In such situations, (15) can be rewritten as

$$\dot{x}^{(k+1)}(t_{SW}) = \frac{2}{h}(x^{(k+1)}(t_{SW}) - x(t_{SW-})) + \dot{x}^{(k)}(t_{SW}) \qquad (24)$$

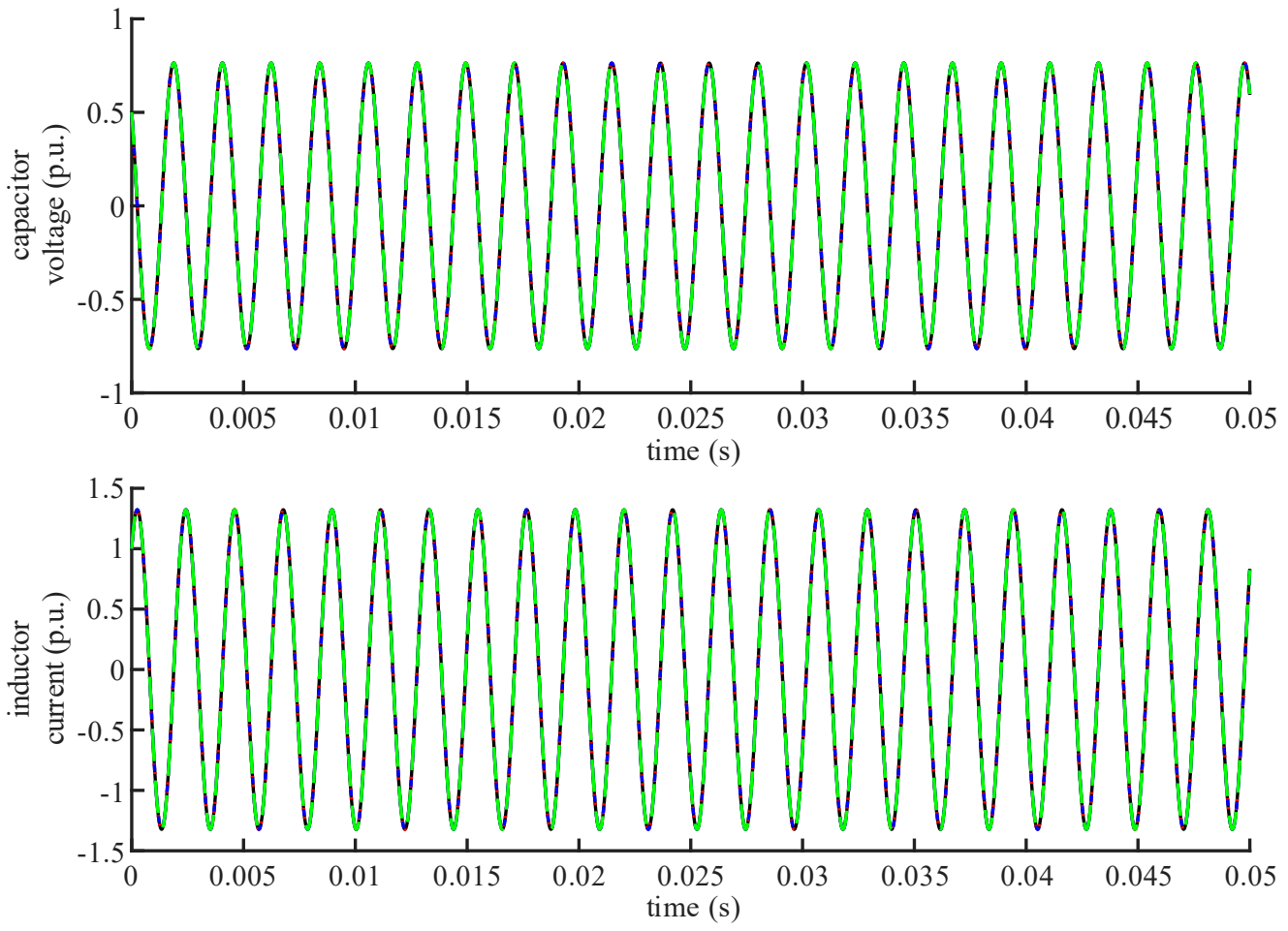

Figure 4. Results from different methods for switing events calculation combined with instantaneous correction compared to the analytical solutions. Solid line: analytical. Dotted line: Method A. Dash-dotted line: Method B. Dashed line: Method C. Numerical results from all the three methods are close to the analytical solution in both subplots.

In EMT simulation, h is usually small. So the 1st term on the right-hand side of (24) becomes a spike due to the instantaneous change in the differential state variable. Overall, the corrected derivative is the initial estimation superimposed with accumulating spikes, which forms a reasonable numerical approximation to the Dirac impulse.

### D. Instantaneous Correction in Reinitialization

In reinitialization, values at $t_{SW}$ have been calculated. Suppose that the derivative of a differential state variable at $t_{SW+}$ has been estimated, (15) can still be used to correct the values with slight modifications: changing $t_{SW}$ to $t_{SW+}$, and $t_{SW-}$ to $t_{SW}$.

Similarly, the convergence analysis still holds with slight modifications in the notation. Differential state variables remain unchanged in reinitialization, because although they can be discontinuous, they are assumed to be right-continuous.

## V. Numerical Examples

### A. Effectiveness of Instantaneous Correction

The circuit with switches in Section II is revisited here. Methods A, B and C combined with instantaneous correction are applied for the simulations, while other settings remain unchanged. Numerical results are compared to the analytical solutions in Fig. 4. With instantaneous correction, all the three methods can produce numerical results that are considerably close to the analytical solutions. Comparing Figs. 2 and 4, the effectiveness of instantaneous correction in mitigating the numerical stability issues is demonstrated.

### B. Improvement in Simulation Accuracy

Although instantaneous correction is proposed aiming at mitigating the numerical stability issues, it helps to improve the accuracy of simulation results as well. The single-phase semi-controlled full bridge rectifier example in [9] is revisited here

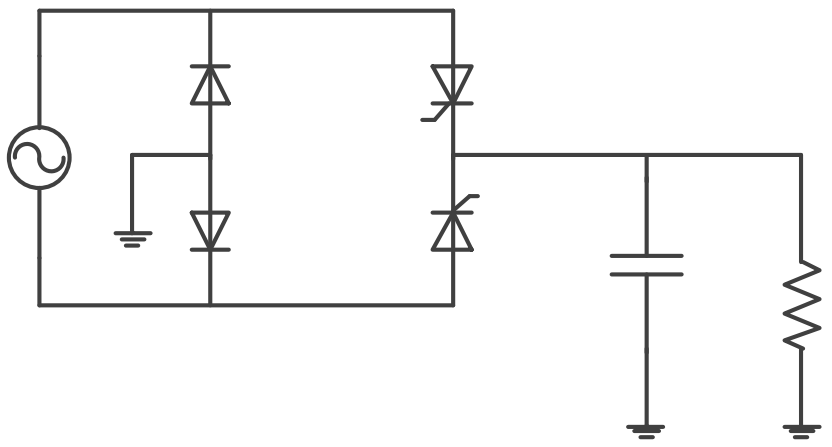

Figure 5. Single-phase semi-controlled full bridge rectifier driving a resistance load [9].

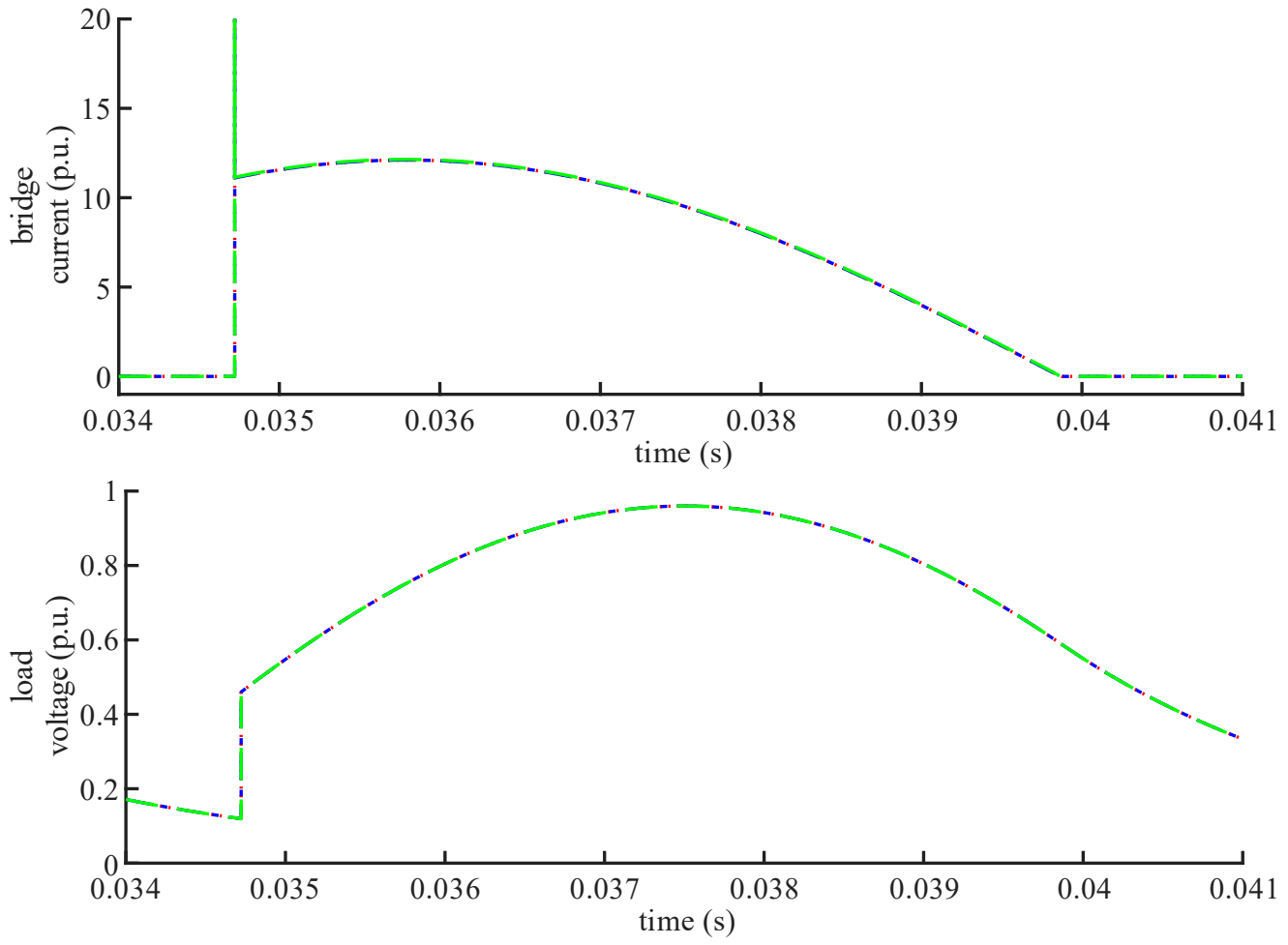

Figure 6. Simulation results from different methods with instantaneous correction. Dotted line: Method A. Dash-dotted line: Method B. Dashed line: Method C. The bridge current spike and the load voltage jump are preciesely represented as vertical lines instead of steep slopes. Results from all the three methods are close. The dash-dotted line and dashed line basically overlap each other in both subplots.

TABLE I. COMPUTED LOAD VOLTAGE DISLOCATION WHEN BRIDGE CURRENT DROPS TO ZERO

| Step Size (µs) | Load Voltage Dislocation | | |
|---|---|---|---|
| | Method | | |
| | A | B | C |
| 10 | 0.0 | 0.0 | 0.0 |
| 20 | 0.0 | 0.0 | 0.0 |
| 50 | 0.0 | 0.0 | 0.0 |
| 100 | 0.0 | 0.0 | 0.0 |
| 200 | 0.0 | 0.0 | 0.0 |
| 500 | 0.0 | 0.0 | 0.0 |

Note: in these simulations, the tolerance for instantaneous correction is $1 \times 10^{-8}$. Numbers smaller than this tolerance are ignored.

as a demonstration. Fig. 5 shows the circuit diagram. Parameters are the same as those in [9].

Fig. 6 presents the bridge current and load voltage generated with the three methods and instantaneous correction, using a 100 $\mu s$ step size for forward time stepping. At around 0.03472 s, a spike appears in the bridge current. This is due to an instantaneous change in the capacitor voltage resulting from the connection between the voltage source and the capacitor through the ideal thyristor, which is turned on by the firing signal. At around 0.03987 s, the bridge current drops to zero turning off the thyristor, while the load voltage is continuous. Intuitively, instantaneous correction improves the simulation accuracy in that the results from all the three methods are reasonably close. Without instantaneous correction, however, results generated with Method A exhibit visible mismatches [9].

As the load voltage is continuous when the bridge current drops to zero, the dislocation in the load voltage between $t_{SW-}$ and $t_{SW+}$ is zero ideally. Thus, the computed load voltage dislocation quantifies the accuracy of a method: the smaller, the more accurate. Table I compares the computed load voltage dislocation of the three methods with different step sizes. In all the considered cases, the dislocation is negligible. Similar comparison without instantaneous correction is given in [9], where the dislocation cannot be ignored. The improved accuracy from instantaneous correction is thus demonstrated.

## VI. CONCLUSION AND FUTURE WORK

Some existing methods for switching events calculation can cause numerical stability issues, either over-stabilizing or destabilizing the studied system. The numerical stability issues result from the property of the methods' transfer functions, which is revealed by the s-domain analysis. The proposed instantaneous correction succeeds in mitigating the numerical stability issues while improving the simulation accuracy.

As a direction for future research, more sophisticated stopping criteria may be investigated balancing accuracy and efficiency.